# Inferring electrochemical performance and parameters of Li-ion batteries based on deep operator networks

Qiang Zheng[1], Xiaoguang Yin[2], Dongxiao Zhang[2,*]

[1]Department of Mathematics and Theories, Peng Cheng Laboratory, Shenzhen 518000, P. R. China

[2]School of Environmental Science and Engineering, Southern University of Science and Technology, Shenzhen 518055, P. R. China

\* Correspondence to: zhangdx@sustech.edu.cn

## Abstract

The Li-ion battery is a complex physicochemical system that generally takes applied current as input and terminal voltage as output. The mappings from current to voltage can be described by several kinds of models, such as accurate but inefficient physics-based models, and efficient but sometimes inaccurate equivalent circuit and black-box models. To realize accuracy and efficiency simultaneously in battery modeling, we propose to build a data-driven surrogate for a battery system while incorporating the underlying physics as constraints. In this work, we innovatively treat the functional mapping from current curve to terminal voltage as a composite of operators, which is approximated by the powerful deep operator network (DeepONet). Its learning capability is firstly verified through a predictive test for Li-ion concentration at two electrodes. In this experiment, the physics-informed DeepONet is found to be more robust than the purely data-driven DeepONet, especially in temporal extrapolation scenarios. A composite surrogate is then constructed for mapping current curve and solid diffusivity to terminal voltage with three operator networks, in which two parallel physics-informed DeepONets are firstly used to predict Li-ion concentration at two electrodes, and then based on their surface values, a DeepONet is built to give terminal voltage predictions. Since the surrogate is differentiable anywhere, it is endowed with the ability to learn from data directly, which was validated by using terminal voltage measurements to estimate input parameters. The proposed surrogate built upon operator networks possesses great potential to be applied in on-board scenarios, such as battery management system, since it integrates efficiency and accuracy by incorporating underlying physics, and also leaves an interface for model refinement through a totally differentiable model structure.

# 1. Introduction

In recent years, increasing numbers of countries and regions globally have promulgated policies to develop and adopt green energy in order to alleviate the crisis from climate change and shortage of fossil fuels. As an energy storage device, Lithium-ion (Li-ion) batteries play an important role in shifting peak load for power systems, and also serve as energy sources for electric vehicles (EVs), since they possess the advantages of high specific energy density, long life, low self-charging rate, and no memory effect [1]. No matter for large energy storage system or battery pack of EVs, a battery management system (BMS) is pivotal to guarantee the safe operation of batteries and full utilization of energy. However, the states used for decisions in BMS are unobservable, e.g., state of charge (SOC) and state of health (SOH), and they need to be connected with observable quantities, e.g., current or voltage, via mathematic models [2].

To describe relationships among different quantities utilized in BMS, several kinds of models are commonly adopted. The models with the highest fidelity are physics-based ones. Most physics-based models originate from the classical pseudo-two-dimensional (P2D) model [3-6], which is built upon porous electrode theory, concentrated solution theory, and Butler-Volmer electrochemical reaction kinetics. Even though the P2D model is based on rigorous theories, solving its coupling systems of partial differential equations (PDEs) is highly CPU-demanding, and inferring a large number of model parameters from measurements is also challenging, which limit its practical applications to BMS. Compared to physics-based models, equivalent circuit models (ECMs) are more efficient since they only describe external characteristics (i.e., current and voltage) of batteries with circuits composed of simple elements, e.g., resistance and capacitance [7]. Therefore, ECMs seem to be more suitable for BMS due to their simple structure and light calculation process [8]. However, due to the lack of physical constraints, ECMs may provide unreliable predictions, which poses a potential risk for battery safety. Besides, analytical models [9,10] and black-box models [11-13] are also used for battery modeling, but they also face the problem of lacking physical insights, even though they are all qualified for efficient evaluation. Therefore, it is required to build a battery model, which is not only fast executable but also physically constrained, for on-board applications in BMS.

Recently, deep learning techniques have been widely used to model complex systems, e.g., hydrology [14,15], biology [16,17], and meteorology [18,19], due to their powerful fitting capabilities, especially in nonlinear and high-dimensional problems. However, deep learning methods usually suffer from a shortage of interpretability and robustness of predictions since they adopt purely data-driven workflows. To address this issue, Raissi et al. [20] proposed physics-informed neural networks (PINNs) to extract information from data and governing equations, which are imposed by automatic differentiation techniques. In such way, the prior information in the form of differential equations can be incorporated by the trained model, whose predictions can thus honor the underlying physics. The proposal of PINNs stimulated the development of scientific machine learning, and contributed to many applications and adaptive improvements in different research domains, e.g., fluid mechanics [21], quantum chemistry [22], material science [23], and geophysics [24]. In the realm of battery modeling, the concept of combining data-driven and physics-based models has attracted more and more attention, with an emphasis on safety management [25], remaining life prediction of batteries [26], etc. More specifically, He et al. [27] proposed a physics-constrained deep neural network to estimate parameters in the zero-dimensional model of the vanadium redox flow battery. Li et al. [28] developed a physics-informed data-driven method to estimate internal concentrations and potentials in electrodes and electrolytes.

The successful applications of deep learning to modeling complex systems are based on the universal approximation theorem, which states that neural networks (NNs) can approximate any continuous function, leading to a mapping from one finite-dimensional space to another one [29]. Besides, there is another less known theorem stating that NNs can be used to approximate any continuous nonlinear functional [30] or operator [31]. Unlike function regression, the operator regression aims to build a mapping between two infinite-dimensional function spaces. Based on the universal approximation for operators, Lu et al. [32] designed a novel network architecture, termed the deep operator network (DeepONet), to learn both explicit operators, e.g., integrals and fractional Laplacians, and implicit operators, e.g., deterministic and stochastic differential equations, from data. Compared to normal NNs built for function regression, DeepONet can take advantage of operator regression and allow for modeling the dynamics of complex systems, e.g., Li-ion batteries, corresponding to different boundary and initial conditions without the need for retraining the networks. Since DeepONet works on function spaces, it is important to draw sufficient training samples to guarantee full exploitation of the spaces, which leads to a heavy computational burden. To ease the massive requirements for data, Wang et al. [33] proposed a physics-informed DeepONet by borrowing the idea from PINNs. Owing to explicit incorporation of governing equations, the robustness of model predictions can also be enhanced. Even though

DeepONet is expressive and built upon rigorous theorem [31], its application is still in its infancy, and no researches have yet been performed focusing on battery modeling using DeepONet.

Although the governing equations for describing dynamical processes in batteries are complex and highly coupled, the battery can still be viewed as a system with a single input and a single output. The input is applied current, and the output is terminal voltage, both of which are functions. Therefore, given a specific battery system, the mapping from input (i.e., current) to output (i.e., voltage) can be understood as a composite operator, which can be represented in various ways, e.g., a system of PDEs, equivalent circuit, or data-driven models. Different ways for demonstrating the operator correspond to models with different levels of accuracy and efficiency, and achieving optimal balances between them remains challenging when developing battery models.

In this work, we aim to build a physics-informed data-driven battery model that reconciles accuracy and efficiency from the perspective of operator learning. Specifically, we firstly compare the performance of DeepONet and physics-informed DeepONet through an operator learning task. We also construct a composite surrogate composed of several DeepONets for mapping from current curve to terminal voltage while considering a material property, i.e., solid diffusivity in this work. In the composite surrogate, Li-ion concentration in solid electrodes plays the role of the intermediate variable, which determines SOC physically, and its predictive accuracy is improved through incorporating underlying physics. More importantly, the composite surrogate built on operator networks is a totally differentiable surrogate system, which means that the input parameters can be estimated based on voltage measurements via direct optimization. Such efficient workflow for inverse modeling cannot be realized by physics-based models, since they treat physical parameters as determined values and do not allow for learning from data, even though some material parameters may vary with the utilization of batteries and need to be updated in time.

The remainder of this paper proceeds as follows. In Section 2, we firstly introduce the physics-based models for simulating electrochemical performance within the Li-ion battery, and then provide a brief introduction of operator learning methods. Section 3 presents the results of operator learning with prior information from governing equations incorporated, and parameter estimation based on composite operator networks. Finally, conclusions and suggestions for future work are given in Section 4.

## 2. Methodology
### 2.1 Simulation of Li-ion battery with extended single particle model

The pseudo-two-dimensional (P2D) model is the most widely used model that describes physics inside of the battery, including the diffusion process of Li-ion in solid and liquid phase, Ohm's law in solid and liquid phase, charge conservation, and Butler-Volmer kinetics [3-6]. However, the P2D model has too many parameters and numerically solving it is expensive, which has contributed to the development of various simplified versions. The single particle (SP) model is one of the simplified models of P2D, and ignores the diffusion of Li-ion along electrode directions [34]. Unfortunately, its underlying assumption of constant Li-ion concentration in electrolyte limits its practical usage of up to 1C current rate [35]. To improve its applicability, the extended single particle (eSP) model has been developed by considering stress-diffusion coupling [36] and electrolyte physics [37,38] based on the SP model. In this work, we take the eSP model as the underlying physical model and use it to generate data for training operator networks.

In the eSP model, the transport of Li-ion is driven by not only concentration gradient, but also mechanical stress field. The diffusion equation coupled with stress-diffusion effect is defined as follows:

$$\frac{\partial c_{s,m}}{\partial t} + \frac{1}{r^2}\frac{\partial}{\partial t}\left[-r^2(1+\Theta_{s,m}c_{s,m})D_{s,m}\frac{\partial}{\partial r}c_{s,m}\right] = 0, \quad m = n, p \tag{1}$$

where $c_{s,m}$ is the Li-ion concentration in the solid phase and $D_{s,m}$ represents the solid diffusion coefficient, i.e., $c_{s,n}, D_{s,n}$ for anode and $c_{s,p}, D_{s,p}$ for cathode; $\Theta_{s,m} = \frac{2\Omega_{s,m}^2 E_{s,m}}{9RT(1-\nu_{s,m})}$ carries the information from stress; $\Omega_{s,m}$ is the partial molar volume; $E_{s,m}$ represents Young's modulus; $R$ is the universal gas constant; $T$ is temperature; and $\nu_{s,m}$ denotes Poisson's ratio. The boundary conditions for equation (1) can be set as:

$$\left.\frac{\partial c_{s,m}}{\partial r}\right|_{r=0} = 0, \quad -(1+\Theta_{s,m}c_{s,m})D_{s,m}\left.\frac{\partial c_{s,m}}{\partial r}\right|_{r=R_m} = j_m^{Li}, \tag{2}$$

where $j_m^{Li}$ represents the pore wall flux of Li-ion at the interface between electrode and electrolyte. It can be given as follows upon the assumption of uniform electrochemical reaction:

$$j_n^{Li} = \frac{-i_{app}R_n}{3F(1-\varepsilon_n)L_n}, \quad j_p^{Li} = \frac{i_{app}R_p}{3F(1-\varepsilon_p)L_p}, \tag{3}$$

where $i_{app}$ denotes the applied current density to the battery; $R_n$ and $R_p$ are solid particle radius for anode and cathode, respectively; $F$ is Faraday constant; and $\varepsilon_m$ and $L_m$ (m = n, p) represent porosity and length of two electrodes, respectively.

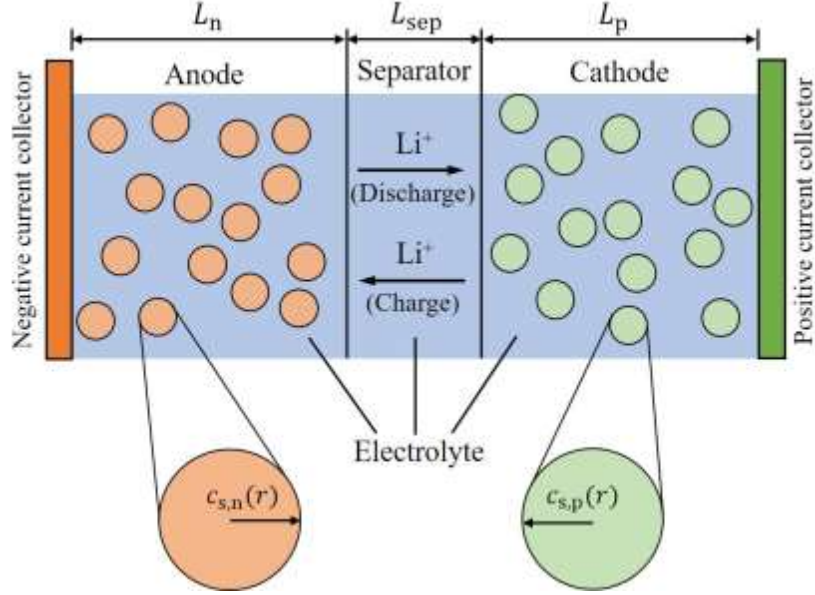

**Fig. 1.** Schematic diagram of Li-ion battery mainly composed of four parts, i.e., anode, separator, cathode, and electrolyte. Current collectors glued to two electrodes are used for collecting electrons.

Apart from taking the diffusion-stress coupling effect into consideration, compared to the single particle (SP) model, the eSP model also considers the effect of electrolyte physics, since the potential drop in electrolyte phase can be an important driver for Li-ion transport. As shown in Fig. 1, the electrolyte phase exists in three parts, i.e., anode, separator and cathode, and Li-ion concentration in these regions can be approximated by polynomial functions [38]:

$$c_{e,n}(x,t) = a_1(1 - e^{-b_1 t})x^2 + a_2(1 - e^{-b_1 t}) + c_{e,0}, \qquad (4)$$

$$c_{e,p}(x,t) = a_3(1 - e^{-b_2 t})(L - x)^2 + a_4(1 - e^{-b_2 t}) + c_{e,0}, \qquad (5)$$

$$c_{e,sep}(x,t) = \frac{(a_3 L_p^2 + a_4)(1 - e^{-b_2 t}) - (a_1 L_n^2 + a_2)(1 - e^{-b_1 t})}{L_{sep}} \times (x - L_n - L_{sep}) \\ + (a_3 L_p^2 + a_4)(1 - e^{-b_2 t}) + c_{e,0}, \qquad (6)$$

where coefficients $a_k, b_k$ ($k = 1, 2, 3, 4$) are further defined in Appendix A; and $c_{e,0}$ represents the initial concentration of Li-ion in electrolyte.

Based on the approximated concentration in the three regions, the electrolyte potential can be given as follows:

$$\phi_{e,n}(x,t) = \phi_{e,n}(0,t) + (1-t_+)\frac{2RT}{F}\ln\frac{c_{e,n}(x,t)}{c_{e,n}(0,t)} - \frac{i_{app}}{2L_n k_{e,n}^{eff}}x^2, \tag{7}$$

$$\phi_{e,sep}(x,t) = \phi_{e,n}(0,t) + (1-t_+)\frac{2RT}{F}\ln\frac{c_{e,sep}(x,t)}{c_{e,n}(0,t)} - \frac{i_{app}}{k_{e,sep}^{eff}}(x-L_n) - \frac{i_{app}L_n}{2k_{e,n}^{eff}}, \tag{8}$$

$$\phi_{e,p}(x,t) = \phi_{e,n}(0,t) + (1-t_+)\frac{2RT}{F}\ln\frac{c_{e,p}(x,t)}{c_{e,n}(0,t)} + \frac{i_{app}}{2L_p k_{e,p}^{eff}}(L-x)^2 \\ -\frac{i_{app}}{2}\left(\frac{L_n}{k_{e,n}^{eff}} + \frac{2L_s}{k_{e,sep}^{eff}} + \frac{L_p}{k_{e,p}^{eff}}\right), \tag{9}$$

where $k_{e,m}^{eff}$ ($m = $ n, p, sep) denotes the effective conductivity of electrolyte in the three regions; and $t_+$ is the Li-ion transference number. As shown in Fig. 1, the length of the battery is $L = L_n + L_{sep} + L_p$. Then, the potential drop of electrolyte across the whole battery can be expressed as:

$$\phi_{e,p}(L,t) - \phi_{e,n}(0,t) = (1-t_+)\frac{2RT}{F}\ln\frac{c_{e,p}(L,t)}{c_{e,n}(0,t)} - \frac{i_{app}}{2}\left(\frac{L_n}{k_{e,n}^{eff}} + \frac{2L_s}{k_{e,sep}^{eff}} + \frac{L_p}{k_{e,p}^{eff}}\right). \tag{10}$$

Based on the Li-ion concentration within the solid electrode particle obtained by equation (1), and potential drop of electrolyte using equation (10), the widely used and most easily measurable quantity, i.e., battery terminal voltage, can be calculated as follows:

$$V_t(t) = \phi_{s,p}(t)\big|_{x=L} - \phi_{s,n}(t)\big|_{x=0} \\ = [U_{OCV,p}(DOD) + \phi_{e,p}(L,t) + \eta_p] - [U_{OCV,n}(SOC) + \phi_{e,n}(0,t) + \eta_n], \tag{11}$$

where $U_{OCV}$ represents open circuit voltage, and it is determined by depth of discharge (DOD) in cathode and state of discharge (SOC) in anode. The DOD and SOC are defined as follows:

$$DOD(t), SOC(t) = \frac{c_{s,m}(R_m,t) - c_{s,m}^{min}}{c_{s,m}^{max} - c_{s,m}^{min}}, \quad m = n, p, \tag{12}$$

where $c_{s,m}(R_m,t)$ denotes Li-ion concentration at the surface ($r = R_m$) of solid electrode particles; $c_{s,m}^{min}$ and $c_{s,m}^{max}$ are the minimum and maximum concentration in solid electrodes, respectively; and $\eta$ represents the overpotential that drives the electrochemical reactions at the solid-electrolyte interface, and it can be described by the Butler-Volmer equation:

$$\mathbf{i} \cdot \mathbf{n} = i_0\left[\exp\left(\frac{\alpha_a F\eta}{RT}\right) - \exp\left(\frac{-\alpha_c F\eta}{RT}\right)\right], \tag{13}$$

where $\mathbf{i}$ is Faradaic current density at the solid-electrolyte interface; $\mathbf{n}$ is the unit normal vector; $i_0$ represents the exchange current density; and $\alpha_a$ and $\alpha_c$ are anodic and cathodic transfer coefficients of electrochemical reactions, respectively.

Even though the battery model is complex, it is still a system with a single input, i.e., applied current $i_{app}$, and a single output, i.e., terminal voltage $V_t(t)$. More importantly, they are usually the only quantities that can be measured in practical use. Based on measurements, we can train a data-driven model, e.g., deep operator networks in this work, with an aim to accelerate prediction in some on-board applications, e.g., BMS. Meanwhile, the existing eSP model can be incorporated into the data-driven model to enhance predictive robustness and this constitutes one of the key points in this work.

## 2.2 Methods for operator learning

### 2.2.1 Deep operator network

Neural networks (NNs) can be used to approximate any continuous function based on the universal approximation theorem. However, another less known theorem also exists, which states that NNs can also approximate continuous nonlinear functional [30] or operators [31], i.e., a mapping from one function space to another function space. Based on the universal approximation theorem for operators, Lu et al. [32] designed a specific network architecture, termed the deep operator network (DeepONet), to learn explicit operators, e.g., integral and fractional Laplacians, as well as implicit operators, e.g., deterministic and stochastic differential equations. A DeepONet is composed of two sub-networks, i.e., a branch net and a trunk net. The branch net takes as input the functions in the form of discretized points, while the trunk net is responsible for encoding the locations where the output functions will be evaluated. Based on different approximation theorems, the branch net can be built in stacked or unstacked forms, and it was shown that the unstacked DeepONet seemed to achieve better performance in terms of generalization error in the original paper [32]. Therefore, in this work, we utilize unstacked DeepONet as the building block.

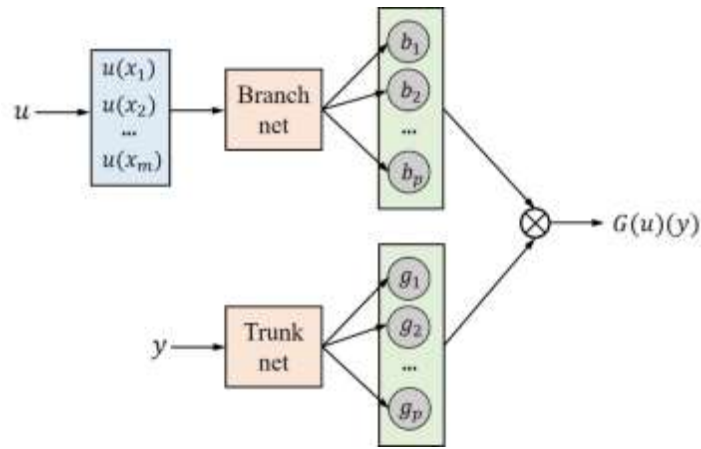

**Fig. 2.** Architecture of the unstacked DeepONet. The branch net takes function $u$ as input, which is discretized at randomly located points $x_1, x_2, \ldots, x_m$. The trunk net takes coordinates $y$ as input, where the output function $G(u)$ can be evaluated. The outputs of branch net and trunk net should share the same size, so that they can conduct a dot production to obtain the final outputs $G(u)(y)$.

As shown in Fig. 2, the unstacked DeepONet aims to learn an operator $G: u \to G(u)$. The branch net is used to encode the input function $u$ at a fixed number of points $x_i, i = 1, 2, \cdots m$, i.e., $[u(x_1), u(x_2), \ldots, u(x_m)]$. It is worth noting that there are no constraints for the location of these points, and they can be chosen randomly or with equal distance. In such condition, the adoption of a fully-connected network structure is appropriate for the branch net. However, if the input function is image-like data and it is discretized with equal-spaced grids, a branch net with a convolutional neural network structure may be more efficient. In other words, the flexibility in choosing evaluation points for functions and structures for sub-networks constitutes a huge advantage of DeepONet compared to other operator learning methods, e.g., Fourier neural operator [39,40]. The trunk net takes the coordinates $y$ as input where the output function can be evaluated. The outputs of branch net $[b_1, b_2, \ldots, b_p]$ and those of trunk net $[g_1, g_2, \ldots, g_p]$ should share the same length, in order to conduct dot production to obtain the final outputs, i.e.:

$$G(u)(y) = \sum_{k=1}^{p} b_k g_k. \tag{14}$$

When training the networks, the data should be organized in the form of triplets, i.e., $(u, y, G(u)(y))$, with the former two representing inputs and the last one representing output. Suppose that we draw $n_1$ separate function samples $u \in \mathcal{U}$, i.e., $\{u^{(i)}\}_{i=1}^{n_1}$, from function space $\mathcal{U}$, and for each input $u^{(i)}$, we collect $n_2$ points, i.e., $\{y_{u,j}^{(i)}\}_{j=1}^{n_2}$, to evaluate the output function. Therefore, the total number of training samples should be $n_1 \times n_2$, and the networks can be trained by adopting the widely used mean square error as loss, i.e.:

$$L_{\text{MSE}} = \frac{1}{n_1 \times n_2} \sum_{i=1}^{n_1} \sum_{j=1}^{n_2} \left[ G(u^{(i)})(y_{u,j}^{(i)}) - \hat{G}(u^{(i)})(y_{u,j}^{(i)}) \right]^2, \tag{15}$$

where $\hat{G}(u^{(i)})(y_{u,j}^{(i)})$ is the predicted value; and $G(u^{(i)})(y_{u,j}^{(i)})$ is the corresponding label data. Since DeepONet is used to learn a mapping between two infinite-dimensional functions, the samples collected to represent function space cannot be very small if the space is rather complex. Consequently, the training dataset with size $n_1 \times n_2$ will be much too large, which will markedly increase the burden of data collection and memory requirements for training. To address this potential issue, physical loss can be incorporated as a regularizer to alleviate the requirements for data, which will be discussed in the next section.

### 2.2.2 Physics-informed deep operator network (PI-DeepONet)

Even though DeepONet is built on rigorous theorem and powerful in operator learning, it still faces two fundamental challenges. Firstly, a quite large corpus of paired data is required to represent the function space and train the networks. In practice, acquisition of such paired data involves a large number of repeated experiments or evaluations of expensive numerical simulators, and thus data preparation may incur a huge cost. Another challenge is that as a purely data-driven method, DeepONet may give the predictions that cannot satisfy the underlying physics, especially when the testing input is outside of the domain of the training dataset. To address these challenges, Wang et al. [33] extended the original DeepONet to a physics-informed version, i.e., PI-DeepONet, inspired by the success of physics-informed neural networks (PINNs). Owing to the differentiable structure and the coordinates taken by the trunk net as inputs, one can easily bias the outputs to satisfy physical constraints by formulating appropriate regularization terms through automatic differentiation techniques. As shown in Fig. 3, the proposed PI-DeepONet differs in the additional loss originated from partial differential equation (PDE), initial condition (IC) and boundary condition (BC), compared with the DeepONet architecture.

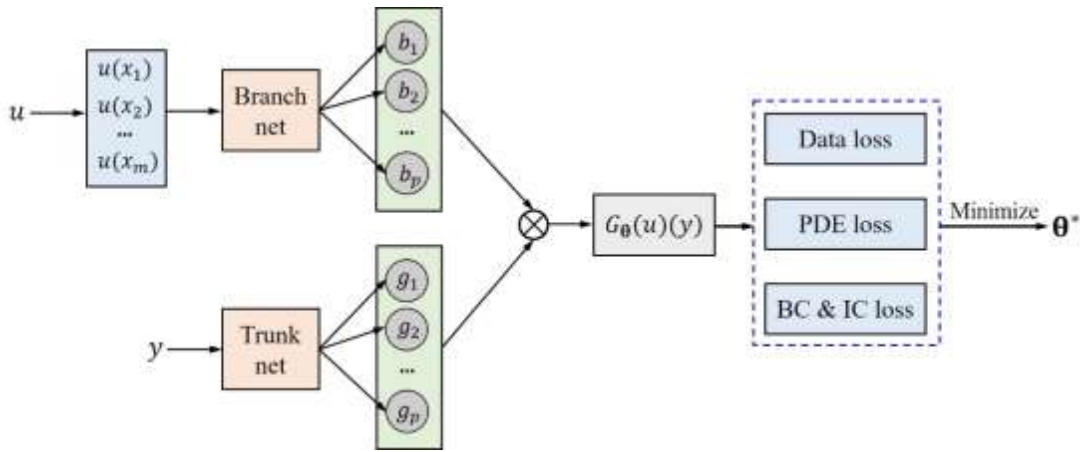

**Fig. 3.** Architecture of the physics-informed DeepONet, i.e., PI-DeepONet, which is parameterized by $\boldsymbol{\theta}$. Based on the output of DeepONet, i.e., $G_{\boldsymbol{\theta}}(u)(y)$, the differential terms can be calculated via automatic differentiation techniques. Therefore, the residuals from partial differential equation (PDE), boundary condition (BC), and initial condition (IC) can be formulated as regularized loss to assist the data loss to guide the training of networks.

In PI-DeepONet, the training data are collected in the same way as in DeepONet. Suppose that we draw $n_1$ separate functions $\{u^{(i)}\}_{i=1}^{n_1}$ from input function space, and for each $u^{(i)}$, there are $n_2$ locations $\{y_{u,j}^{(i)}\}_{j=1}^{n_2}$ being determined by data observations. Such data will be used to build data loss with equation (15). Meanwhile, in order to explore function space more efficiently, we draw different samples of input function and locations of output function for enforcing the IC, BC and PDE constraints, with $\{u^{(i)}\}_{i=1}^{n_{1,\text{IC}}}$ and $\{y_{u,j}^{(i)}\}_{j=1}^{n_{2,\text{IC}}}$ for IC loss, $\{u^{(i)}\}_{i=1}^{n_{1,\text{BC}}}$ and $\{y_{u,j}^{(i)}\}_{j=1}^{n_{2,\text{BC}}}$ for BC loss, and $\{u^{(i)}\}_{i=1}^{n_{1,\text{PDE}}}$ and $\{y_{u,j}^{(i)}\}_{j=1}^{n_{2,\text{PDE}}}$ for PDE loss. Consider a general parametric PDE taking the following form:

$$\mathcal{N}(u, G(u)) = 0, \tag{16}$$

where $u$ represents the parameters (i.e., input functions); $G(u)$ denotes the corresponding solutions of PDE; and $\mathcal{N}$ is a linear or nonlinear differential operator. Therefore, the training loss from different sources can be calculated as follows:

$$L_{\text{data}} = \frac{1}{n_1 \times n_2} \sum_{i=1}^{n_1} \sum_{j=1}^{n_2} \left[ G(u^{(i)})(y_{u,j}^{(i)}) - \hat{G}(u^{(i)})(y_{u,j}^{(i)}) \right]^2, \tag{17}$$

$$L_{\text{IC}} = \frac{1}{n_{1,\text{IC}} \times n_{2,\text{IC}}} \sum_{i=1}^{n_{1,\text{IC}}} \sum_{j=1}^{n_{2,\text{IC}}} \left[ G(u^{(i)})(y_{u,j}^{(i)}) - \hat{G}(u^{(i)})(y_{u,j}^{(i)}) \right]^2, \tag{18}$$

$$L_{\text{BC}} = \frac{1}{n_{1,\text{BC}} \times n_{2,\text{BC}}} \sum_{i=1}^{n_{1,\text{BC}}} \sum_{j=1}^{n_{2,\text{BC}}} \left[ G(u^{(i)})(y_{u,j}^{(i)}) - \hat{G}(u^{(i)})(y_{u,j}^{(i)}) \right]^2, \tag{19}$$

$$L_{\text{PDE}} = \frac{1}{n_{1,\text{PDE}} \times n_{2,\text{PDE}}} \sum_{i=1}^{n_{1,\text{PDE}}} \sum_{j=1}^{n_{2,\text{PDE}}} \left[ \mathcal{N}\left(u^{(i)}, \hat{G}(u^{(i)})(y_{u,j}^{(i)})\right) \right]^2, \tag{20}$$

$$L_{\text{total}} = \lambda_{\text{data}} L_{\text{data}} + \lambda_{\text{IC}} L_{\text{IC}} + \lambda_{\text{BC}} L_{\text{BC}} + \lambda_{\text{PDE}} L_{\text{PDE}}, \tag{21}$$

where $\hat{G}(u^{(i)})(y_{u,j}^{(i)})$ are predicted values, while $G(u^{(i)})(y_{u,j}^{(i)})$ represent the corresponding labels; and the coefficients $\lambda_{\text{data}}$, $\lambda_{\text{IC}}$, $\lambda_{\text{BC}}$, and $\lambda_{\text{PDE}}$ are weights used to adjust the relative importance of each item. In this work, we applied PI-DeepONet to learn an operator for mapping applied current density (i.e., $i_{\text{app}}$) to Li-ion concentration in solid phase (i.e., $c_{\text{s,m}}$) by incorporating the prior physical constraints represented by equations (1)-(3). For additional details about PI-DeepONet, one can refer to [33].

## 3. Results

### 3.1 Operator learning for Li-ion concentration in solid phase

In this section, we build a predictive model for Li-ion concentration in solid electrodes based on input current functions, with an aim to test the learning capability of DeepONet and determine whether PI-DeepONet can outperform DeepONet. Since Li-ion concentration cannot be measured, we adopt the eSP model working in different applied currents (i.e., $i_{app}$) to generate training and testing datasets. In this work, we consider discharge scenarios with time-varying currents, which are set as $i_{app}(t) = i_{1C} \times N(t)$, during a period of 3600 s. The 1C rate is taken as $i_{1C} = 27$ A/m$^2$ according to battery geometry and specific capacity of electrodes. The discharge rate $N(t)$ varies with time, and is defined as follows:

$$h(t) = \begin{cases} \frac{1}{1800}t, & 0 < t \leq 1800 \\ -\frac{1}{1800}t + 2, & 1800 < t \leq 3600 \end{cases}, \tag{22}$$

$$N(t) = -h(t) * \beta + 1, \tag{23}$$

where $\beta$ is a key coefficient to account for the varieties of current functions. The eSP model is built with COMSOL Multiphysics, and the parameters for model creation are shown in Table 1. In practice, we run the eSP model repeatedly by changing the input current, and collect each corresponding Li-ion concentration in solid phase as model outputs. Since the current function $i_{app}$ is totally determined by discharge rate function $N$ in this work, we adopt function $N$ as model inputs. Based on the paired data of inputs and outputs, the operator networks can be trained and tested in different ways, e.g., interpolation and extrapolation.

**Table 1.** Parameters used in the eSP model.

| Parameters | Symbol | Value | Unit |
|---|---|---|---|
| **General** | | | |
| Universal gas constant | $R$ | 8.314 | J/mol/K |
| Temperature | $T$ | 298 | K |
| Initial Li-ion concentration in electrolyte | $c_{e,0}$ | 2000 | m$^3$/mol |
| Anodic symmetry factor | $\alpha_a$ | 0.5 | - |
| Cathodic symmetry factor | $\alpha_c$ | 0.5 | - |

| Separator | | | |
|---|---|---|---|
| Porosity | $\varepsilon_{\text{sep}}$ | 0.42 | - |
| Length | $L_{\text{sep}}$ | 20 | $\mu m$ |
| Effective diffusion coefficient of electrolyte | $D_{e,\text{sep}}^{\text{eff}}$ | $3 \times 10^{-10} \times \varepsilon_{\text{sep}}^{1.5}$ | m²/s |
| Effective electrical conductivity of electrolyte | $k_{e,\text{sep}}^{\text{eff}}$ | $3 \times \varepsilon_{\text{sep}}^{2.3}$ | S/m |

| Electrodes | | Anode (graphite) | Cathode (LiCoO$_2$) | |
|---|---|---|---|---|
| Porosity | $\varepsilon_{\text{m}}$ | 0.35 | 0.35 | - |
| Length | $L_{\text{m}}$ | 70 | 70 | $\mu m$ |
| Radius of solid particle | $R_{s,\text{m}}$ | 12 | 12 | $\mu m$ |
| Maximum Li-ion concentration in particle | $c_{s,\text{m}}^{\max}$ | 31250 | 49500 | m³/mol |
| Exchange current density | $i_{0,\text{m}}$ | 36 | 26 | A/m² |
| Diffusion coefficient of solid phase | $D_{s,\text{m}}$ | $4 \times 10^{-14}$ | $1 \times 10^{-13}$ | m²/s |
| Partial molar volume | $\Omega_{\text{m}}$ | $3.1 \times 10^{-6}$ | $-7.28 \times 10^{-7}$ | m³/mol |
| Young's modulus | $E_{\text{m}}$ | 15 | 375 | GPa |
| Poisson's ratio | $\nu_{\text{m}}$ | 0.3 | 0.2 | - |
| Effective diffusion coefficient of electrolyte | $D_{e,\text{m}}^{\text{eff}}$ | $3 \times 10^{-10} \times \varepsilon_{\text{n}}^{1.5}$ | $3 \times 10^{-10} \times \varepsilon_{\text{p}}^{1.5}$ | m²/s |
| Effective electrical conductivity of electrolyte | $k_{e,\text{m}}^{\text{eff}}$ | $3 \times \varepsilon_{\text{n}}^{4.1}$ | $3 \times \varepsilon_{\text{p}}^{4.1}$ | S/m |
| Effective electrical conductivity of solid phase | $k_{s,\text{m}}^{\text{eff}}$ | $100 \times \varepsilon_{\text{n}}^{1.5}$ | $10 \times \varepsilon_{\text{p}}^{1.5}$ | S/m |

**Note:** The symbol subscript "m" changes to "n" when representing anode, and "p" when representing cathode.

### 3.1.1 Inference on current function

Firstly, we start from testing the capability of DeepONet in inference on unseen current functions. As shown in Fig. 4, we adopt the unstacked DeepONet proposed in [32] as the basic network architecture. The branch net takes as input the discharge rate function $N(\cdot)$, which are discretized at 101 points (i.e., $[N(t_1), N(t_2), ..., N(t_{101})]$) equally located at the period from 0 to 3600 s in this work. The trunk net takes spatial-temporal coordinates (i.e., $r$ and $t$) as inputs, and the discharge rate values with respect to specific times through equation (23) also serve as the input (i.e., $N(t)$) since they need to be differentiated when building PI-DeepONet. The fully-connected neural network with six hidden layers and each layer containing 200 neurons, is applied to construct the branch net and the trunk net, both of which take softplus as the activation function

[41]. The only difference between branch and trunk nets is the number of neurons in the input layer, with 101 for branch net and 3 for trunk net. In this experiment, we draw nine current functions, i.e., discharge rate functions, with $\beta = 0.1, 0.2, \ldots, 0.9$ for building the training dataset. For each current function, there are 21 spatial points equally located around the radius (i.e., $R_s = 12\mu m$) of electrodes, and 451 temporal points equally located during the simulation period (i.e., 3600 s). The input function, coordinates, and output Li-ion concentration can be organized as training data in triplet form, i.e., $\{[N(t_1), N(t_2), \ldots, N(t_{101})], [r, t, N(t)], [c_s(N(\cdot), r, t)]\}$, and the total number of training samples is 9×21×451=85,239. The Adam optimizer [42] with a learning rate of 5e-4 is employed to train DeepONet with MSE loss. It takes approximately 28 min to train DeepONet on one Tesla T4 GPU for 1,500 epochs.

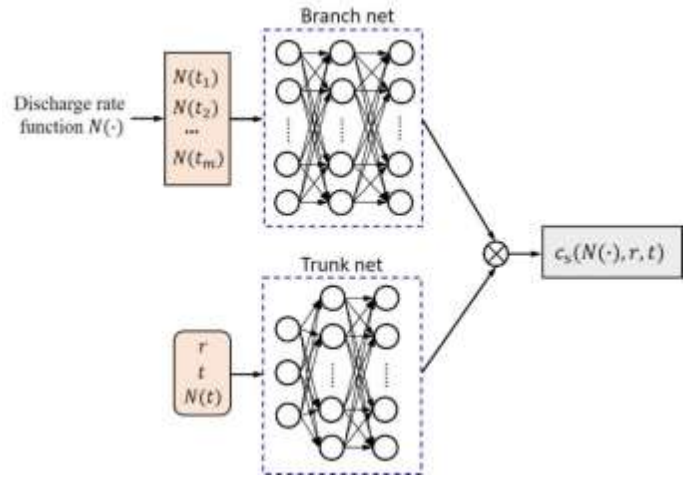

**Fig. 4.** DeepONet architecture for learning the operator mapping from current curve to Li-ion concentration in solid phase.

We train two DeepONets using data from anode and cathode, respectively. After training, we can test its predictive capability in unseen current functions. To achieve this, we set $\beta$=0.25, 0.55 and 0.85 to produce three current curves, and use the eSP model to obtain reference values. As shown in Fig. 5, the matching performance of anode (i.e., Fig. 5(b)-5(d)) and cathode (i.e., Fig. 5(e)-5(g)) under newly different settings of input current, which are denoted by dashed lines in Fig. 5(a), are markedly perfect.

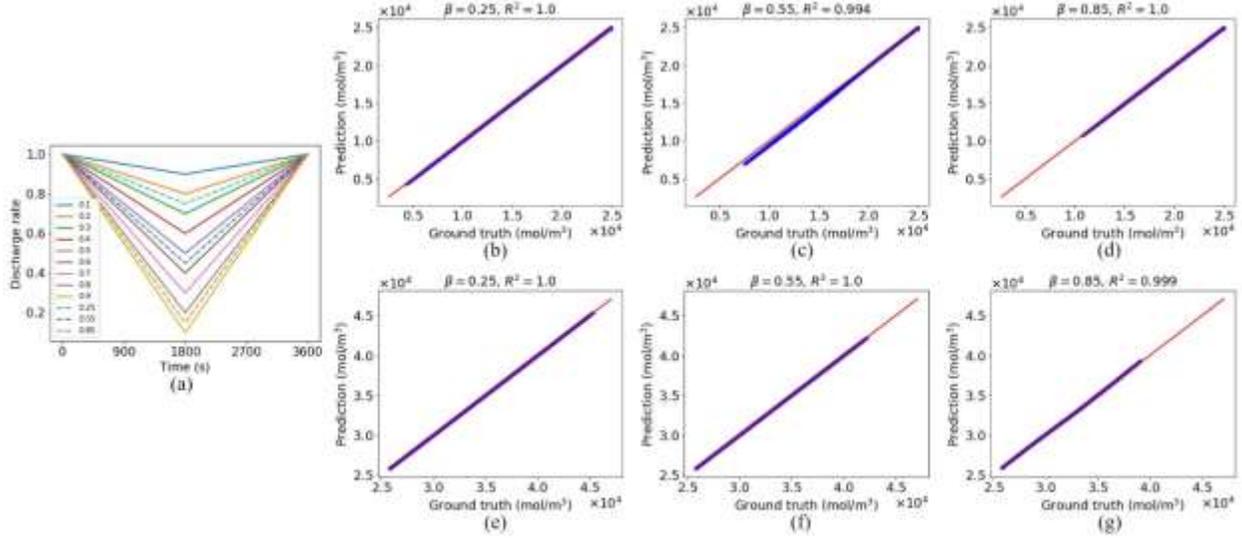

**Fig. 5.** (a) The time-varying discharge rate function $N(t)$ with $\beta$ ranging from 0.1 to 0.9, in which the data with respect to solid lines are used for training and those of dashed lines are for testing. The remaining subfigures demonstrate the testing performance for predicting Li-ion concentrations under three new $N(t)$ functions, denoted by three dashed lines in (a), with $\beta$ being 0.25, 0.55 and 0.85, respectively. Subfigures (b)-(d) represent the performance for anode, while (e)-(g) for cathode.

The above experiment can be viewed as an interpolation test, and we subsequently perform an extrapolation test on current functions. To achieve this, we continue to use data prepared in the last experiment, and set data corresponding to current curves ($\beta$ ranging from 0.2 to 0.8) for training, and the remaining for testing, as shown by solid and dashed lines, respectively, in Fig. 6(a). In addition, we prepare data based on partial differential equation (PDE), initial condition (IC), and boundary condition (BC) to build a PI-DeepONet, with an aim to explore its potential advantages in extrapolated scenarios. The initial concentrations for two electrodes are $c_{s,n}^0 = 25000 \text{ mol/m}^3$ and $c_{s,p}^0 = 25800 \text{ mol/m}^3$, which will serve as the labels to build IC loss using equation (18). Considering that the operator is built for mapping from current to Li-ion concentration of electrodes in this section, the diffusion equation (i.e., equation (1)) is adopted for PDE loss, while its boundary conditions (i.e., equations (2) and (3)) are used for BC loss. The BC and PDE losses are calculated by equation (19) and (20), respectively.

In terms of data preparation for obtaining the above losses, as suggested in section 2.2.2, we draw new samples of current function to construct each item of IC, BC and PDE losses, apart from those used for constructing data loss (i.e., equation (17)). Therefore, we set the range of $\beta$ as 0 to 1, and randomly select 100 samples to produce current functions using equation (23). Meanwhile, for each function, we randomly draw 100 samples of spatial and temporal coordinates within their ranges, i.e., 0-12 $\mu m$ for spatial locations and 0-3600 s for temporal locations, to encode IC loss, and 500 samples to encode BC and PDE losses. As a result, the number of data for building IC loss is 10,000, and that for building BC and PDE losses are both 50,000. The total loss for training PI-DeepONet is organized with equation (21), in which the coefficients for balancing each item are taken as $\lambda_{\text{data}} = 10^3$, $\lambda_{\text{IC}} = 10^4$, $\lambda_{\text{BC}} = 10^2$, and $\lambda_{\text{PDE}} = 10^{-7}$, and meanwhile the data loss (i.e., equation (17)) is used for training DeepONet. The settings of training are the same as those in the last experiment, except for adding the training epochs of PI-DeepONet to 2,500, with an aim to incorporate physical information more sufficiently. Consequently, the training of PI-DeepONet increases to 50 min on one Tesla T4 card.

Upon obtaining the trained models, we employ them to give predictions on unseen current functions, corresponding to $\beta = 0.1, 0.9$, as shown by the dashed lines in Fig. 6(a). The matching performances of Li-ion concentration in anode are shown in Fig. 6(b)-6(e), with (b)(c) representing the results of DeepONet and (d)(e) representing those of PI-DeepONet. The results show that PI-DeepONet has a slight advantage over DeepONet, and the comparisons in cathode are almost the same, and thus they are not displayed here in order to avoid redundancy. There are two possible reasons to account for such a weak advantage. Firstly, DeepONet is powerful in extracting information from function space defined by equation (23). Secondly, the current space used in this work is relatively simple, and testing samples outside of the range of the training dataset are quite similar to the training ones, leading to a relatively weak extrapolation scenario. Therefore, a stronger extrapolation test needs to be considered to elucidate the advantages of PI-DeepONet.

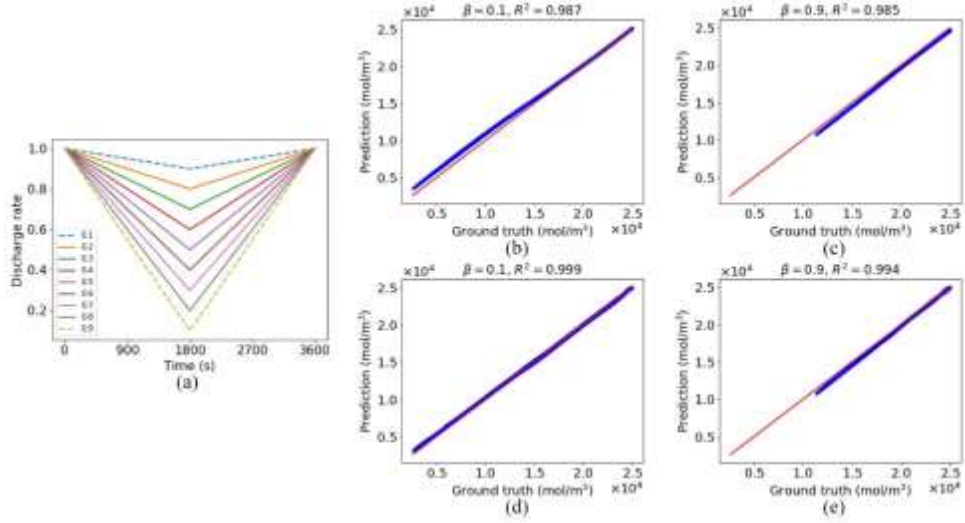

**Fig. 6.** Comparisons of predictive performance between DeepONet and PI-DeepONet for Li-ion concentration in anode. Subfigure (a) denotes the discharge rate functions, with two dashed lines ($\beta$=0.1, $\beta$=0.9) corresponding to testing data and the others serving as training data. The concentration matching performance of DeepONet is illustrated in (b) and (c) for $\beta$=0.1 and $\beta$=0.9, respectively. The matching performance of PI-DeepONet is illustrated in (d) and (e).

### 3.1.2 Temporal extrapolation

To more thoroughly explore the effects of incorporated physics in PI-DeepONet, we consider a temporal extrapolation test in this part. We continue to use the networks and paired data prepared above, i.e., nine current functions with $\beta$ ranging from 0.1 to 0.9 and the corresponding Li-ion concentrations in two electrodes. Based on this dataset, i.e., in the form of $\{[N(t_1), N(t_2), \ldots, N(t_{101})], [r, t, N(t)], [c_s(N(\cdot), r, t)]\}$, we assign those in the first half of the period ($t \leq 1800$ s) as the training dataset, and the remaining ($t > 1800$ s) as the testing dataset. Compared to the previous experiment, the main difference lies in the inputs for the trunk net, with $t$ less than or equal to 1,800 for training and larger than 1,800 for testing. The branch net also takes the entire current function as input, since it serves as a known boundary condition for discharging in this work. The above paired data are adopted to train DeepONet, and also build data loss in PI-DeepONet. Furthermore, the additional data for encoding IC, BC, and PDE loss in the last experiment are also utilized here to train PI-DeepONet. The training settings remain the same as previously, except for modifying the training epochs for DeepONet as 3,000 and that for PI-DeepONet as 5,000.

After training, we can feed the model with the samples in the latter half period to obtain predictions. We select four current functions with $\beta$=0.3, 0.5, 0.7, and 0.9 to demonstrate the predictive performance in anode and cathode. The matching results of DeepONet are shown in Fig. 7(a-b). It can be clearly seen that predictive accuracy becomes increasingly worse as $\beta$ becomes larger in both of the two electrodes. A possible reason for this is that the input current becomes more nonlinear when $\beta$ grows, as shown in Fig. 5(a), and the nonlinearity will be transferred to the relationship between output concentration and input coordinates. It is challenging for the purely data-driven DeepONet to capture the nonlinearity if it is not covered by the training data. In comparison, as shown in Fig. 7(c-d), the results of PI-DeepONet are satisfactory as $\beta$ grows. This is because the physical information in the form of IC, BC and PDE losses acts on the entire temporal domain, and can bias predictions in the testing period to satisfy physical constraints.

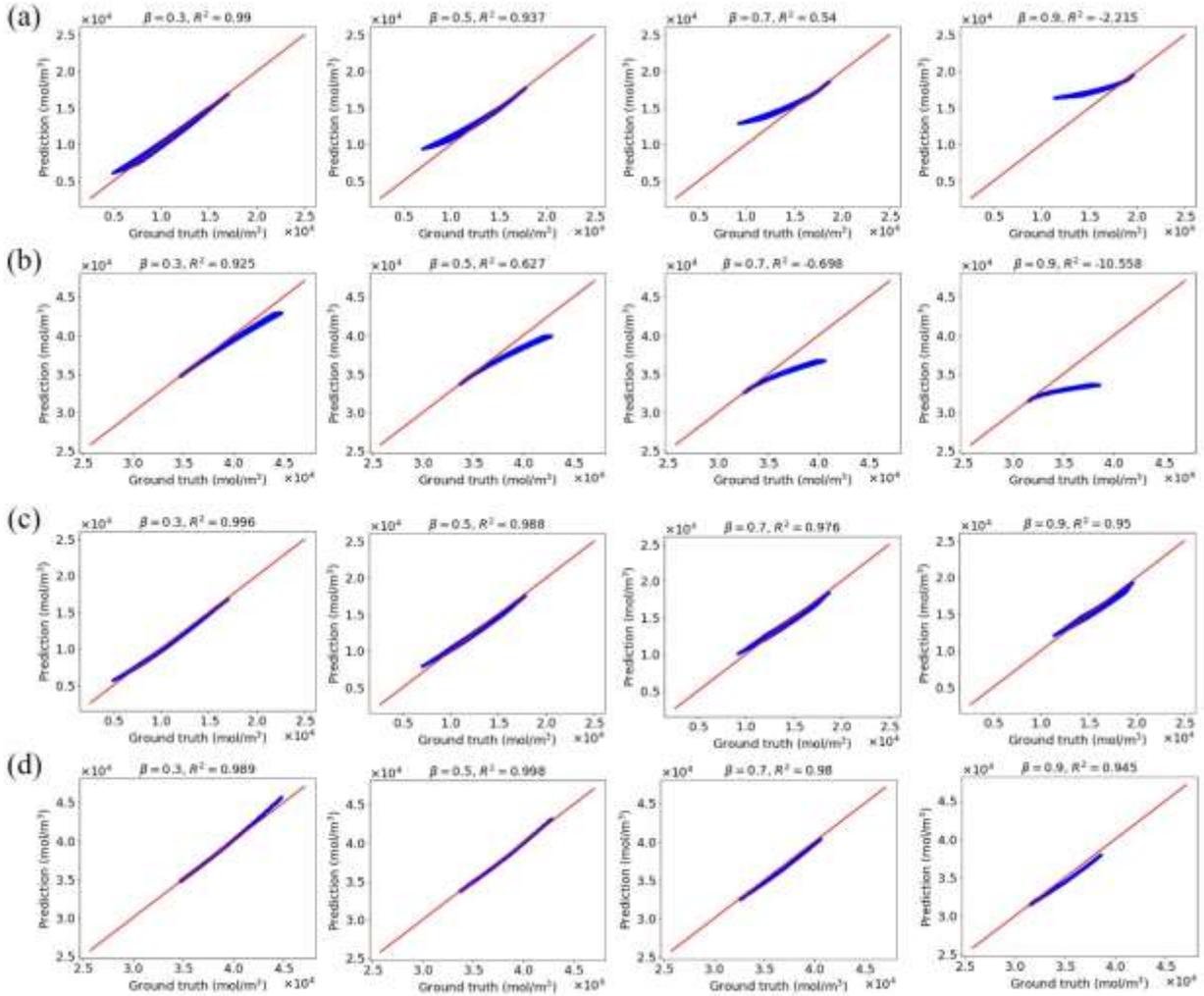

**Fig. 7.** Predictions of Li-ion concentration versus their ground truth in the testing period using (a-b) DeepONet and (c-d) PI-DeepONet. Subfigures (a) and (c) represent comparisons in anode, while (b) and (d) denote the results in cathode. Different subfigures in each column correspond to different $\beta$ values.

In practice, Li-ion concentration in solid electrodes is hardly measurable, and it just serves as an intermediate value to obtain SOC and DOD, and finally obtain terminal voltage predictions according to equation (11). During this process, the relationship between open circuit voltage and SOC or DOD is determined through interpolation, and the concentration value needs to be scaled when calculating SOC or DOD using equation (12). In other words, the inaccuracy in concentration predictions may not pose significantly negative effects for the voltage. To evaluate such effects quantitatively, we calculate terminal voltage through the eSP model based on the above concentration predictions at two electrodes. As shown in Fig. 8, in the training period (i.e., $t \leq 1800$ s), the matching performance of terminal voltage of DeepONet is similar to that of PI-DeepONet. In the testing period (i.e., $t > 1800$ s), however, the predicted terminal voltages in DeepONet are increasingly far from the references as time and $\beta$ increase. In comparison, the predictions of PI-DeepONet still agree well with the references in the testing period, with the largest absolute errors being less than 5 mV. Therefore, through the experiments in this section, it is confirmed that DeepONet is powerful in learning the operator mapping from current to Li-ion concentration, and PI-DeepONet can outperform DeepONet, especially in temporal extrapolation scenarios.

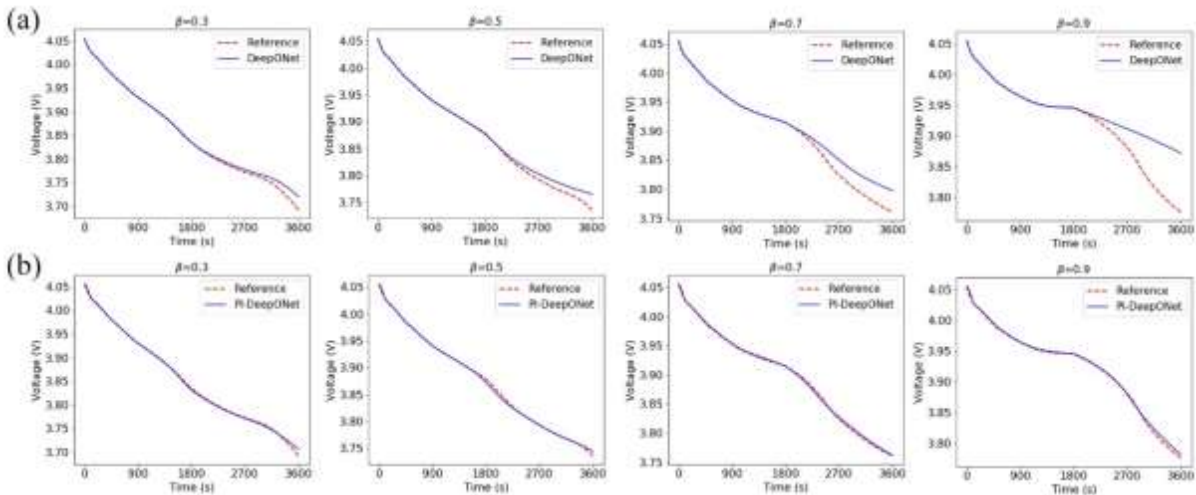

**Fig. 8.** The matching of predicted terminal voltages and their references using (a) DeepONet and (b) PI-DeepONet. The settings of $\beta$ for each column remain consistent with those in Fig. 7.

## 3.2 Surrogate modeling with operator networks by incorporating solid diffusivity

The battery system with specified current and material property as inputs and terminal voltage as output can be described by a composite of operators, which are suitable for surrogate modeling via operator networks, with an aim to improve computational efficiency in practical applications. Therefore, after validating the learning capability of DeepONet and PI-DeepONet in the last section, we proceed in this section to build a surrogate for complex systems by incorporating material property, i.e., diffusion coefficients of solid electrodes in this work, as model inputs. Since Li-ion concentration in electrodes is a key factor to determine SOC, which is needed in BMS, the composite operators are split into two consecutive parts for learning by taking Li-ion concentration as an intermediate quantity. The following two subsections are used to build operators for predicting Li-ion concentration and terminal voltage, respectively.

### 3.2.1 Predicting Li-ion concentration from current curve and solid diffusivity

In this section, we construct an operator for mapping current and solid diffusion coefficient to Li-ion concentration at electrodes. To accomplish this, we set nine diffusion coefficients with uniform spacing, i.e., 2e-14, 3e-14, …, 10e-14, for both anode and cathode. For each diffusion coefficient, we conduct numerical experiments as was done in the last section, i.e., considering nine input current functions with respect to $\beta$=0.1, …, 0.9, and collect the generated data for training. The data are still organized in triplet form as in previous experiments. In order to incorporate information from solid diffusivity, the scalar diffusion coefficients are formulated to a vector with equal values, so as to concatenate with the current function to jointly serve as inputs of the branch net, as shown in Fig. 9. Here, the length of the diffusion coefficient vector is set as 10, and the current function is still discretized at 101 points as previously, and thus the number of input neurons for the branch net should be set as 111. Since PI-DeepONet has been validated in predicting Li-ion concentration with good accuracy, we adopt it here to build a surrogate, and the surrogates for anode and cathode are constructed, respectively. Apart from the input layers of the branch net, the other network settings, e.g., the number of layers, neurons in each layer and activation functions, and training settings, e.g., optimizer, learning rate and training epochs, remain consistent with those used in the temporal extrapolation test.

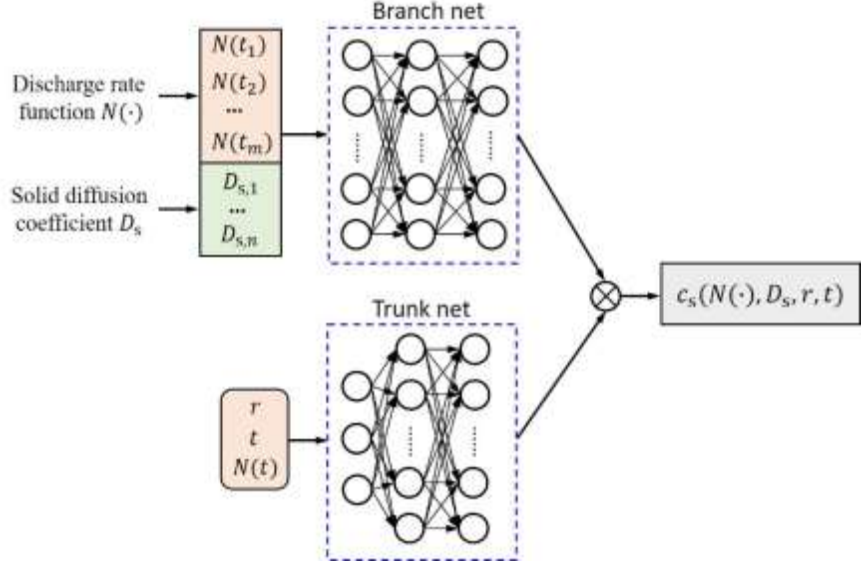

**Fig. 9.** The architecture of PI-DeepONet for predicting Li-ion concentration based on current function, which is concatenated with a vector of solid diffusion coefficients with $n$ equivalent values.

In order to test the predictive accuracy of the trained surrogates, we consider two scenarios, with inputs not seen by the training dataset. In scenario 1, we set $\beta=0.45$ for current function, $D_s^a$=7.5e-4 for diffusion coefficient in anode, and $D_s^c$=6.5e-4 for diffusion coefficient in cathode; whereas, in scenario 2, the above parameters are determined as $\beta=0.65$, $D_s^a$=9.5e-4, and $D_s^c$=4.5e-4, respectively. These parameters are firstly fed to the eSP model to obtain the references for Li-ion concentration in solid phase. Then, the predictions at anode and cathode across the whole spatial and temporal domains can be acquired efficiently with the trained surrogates. The comparisons of predictions and references at two electrodes in two scenarios are given in Fig. 10, in which subfigures (a-b) correspond to the results of scenario 1 and subfigures (c-d) correspond to those of scenario 2. It can be seen from subfigures (a) and (c) that the predictions close to the end time in anode have relatively large deviations. While in cathode, the predictions near two boundaries may pose larger errors, as shown in subfigures (b) and (d). Even though the predictions near the spatial or temporal margins are relatively less accurate, the biggest relative errors in the two scenarios are only approximately 2%, which still indicates satisfactory fitting performance.

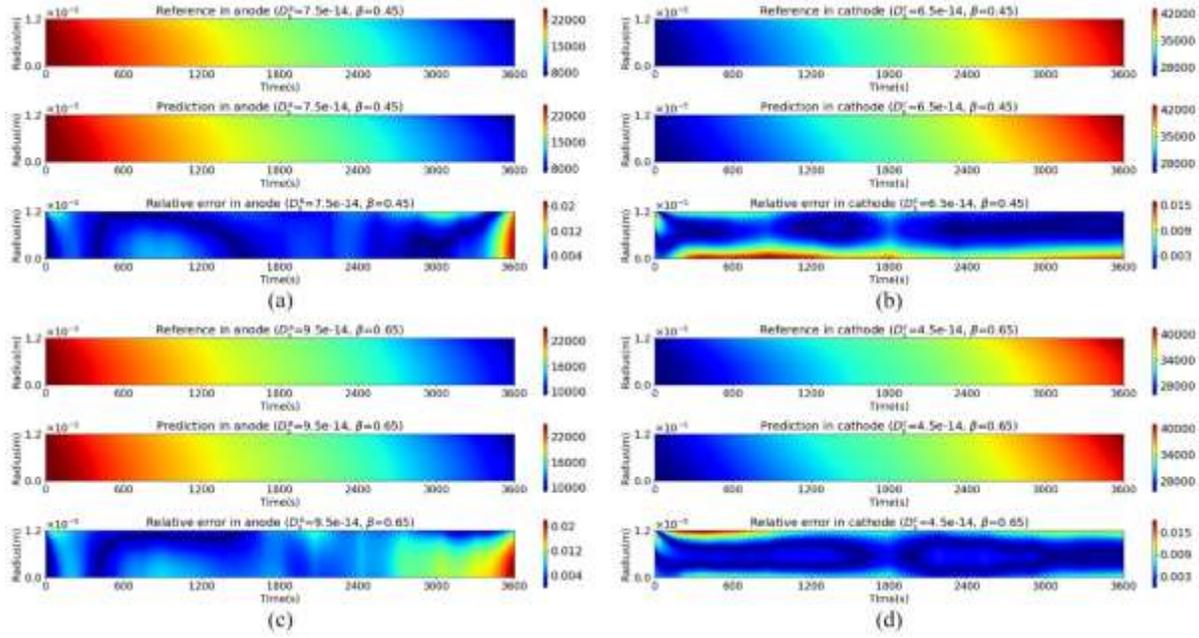

**Fig. 10.** 2D plot of referential and predicted Li-ion concentration, and their relative errors. Subfigures (a-b) correspond to the testing results of scenario 1, while (c-d) are for scenario 2. They differ in the input parameters, i.e., $\beta$ that determines the current curve, diffusion coefficient in anode $D_s^a$, and diffusion coefficient in cathode $D_s^c$. Subfigures (a) and (c) represent the results of anode, while (b) and (d) represent those of cathode.

### 3.2.2 Predicting terminal voltage from surface Li-ion concentration and solid diffusivity

After obtaining the predictions of Li-ion concentration, we can proceed to calculate the terminal voltage, which can be measured in practice, and adopted to examine and refine the predictions. In this work, the terminal voltage is obtained through the eSP model with equation (11), in which the potential drop of electrolyte and overpotential at the surface of electrode particles are analytically calculated by equation (10) and (13), respectively. The predictions of open circuit voltage (OCV) are obtained through interpolation based on the OCV-SOC(DOD) curve, which highly depends on the material of electrodes, i.e., $LiCoO_2$ for cathode and graphite for anode in this work. Even though the terminal voltage needs to be calculated by complex equations, it actually describes the potential drop at the two ends of solid electrodes, which are greatly affected by Li-ion concentration at the surface of electrodes. Consequently, the mapping from surface Li-ion concentration to voltage can be replaced with an operator network as an efficient surrogate.

Since the eSP model employs interpolation when calculating OCV, it cannot be encoded via automatic differentiation techniques, and consequently we adopt DeepONet in this section to build a surrogate.

The schematic of DeepONet for mapping from surface Li-ion concentration to terminal voltage is shown in Fig. 11. Similar to previously, we concatenate the vector filled with equal values of diffusion coefficients to the discretized function of surface concentration, jointly serving as the input of the branch net. The trunk net only needs temporal location and its corresponding discharge rate as inputs and does not require spatial coordinates, since the output terminal voltage denotes the potential drop at two ends. Based on the generated data from the numerical experiments in the last subsection, we extract surface Li-ion concentration at anode and cathode, and the corresponding terminal voltage as the basic training dataset. To incorporate information of surface concentration from two electrodes simultaneously, we make an average of them for branch inputs, i.e., $c_{\text{surf}} = (c_{\text{surf}}^{\text{a}} + c_{\text{surf}}^{\text{c}})/2$, and discretize it at 301 points at an interval of 12 s. Similarly, the vectors of diffusion coefficients of anode and cathode should concatenate with surface concentration, and the length of each vector is set as 10. Eventually, the input for the branch net is composed as $[c_{\text{surf}}(t_1), \cdots, c_{\text{surf}}(t_{301}), D_{\text{s},1}^{\text{a}}, \cdots, D_{\text{s},10}^{\text{a}}, D_{\text{s},1}^{\text{c}}, \cdots, D_{\text{s},10}^{\text{c}}]$, and consequently the dimensionality of the input layer for the branch net should be 321. The terminal voltage for training is selected with an equal interval of 8 s from the entire simulation period, and the paired data are organized in triplet form as previously. Both the branch and trunk nets have four hidden layers with each layer containing 300 neurons, and the rectified linear unit (ReLU) function is used as an activation function for the two networks. The Adam optimizer with a learning rate of 5e-4 is employed to train the networks. It takes approximately 30 min to train 2,000 epochs on one Tesla T4 card.

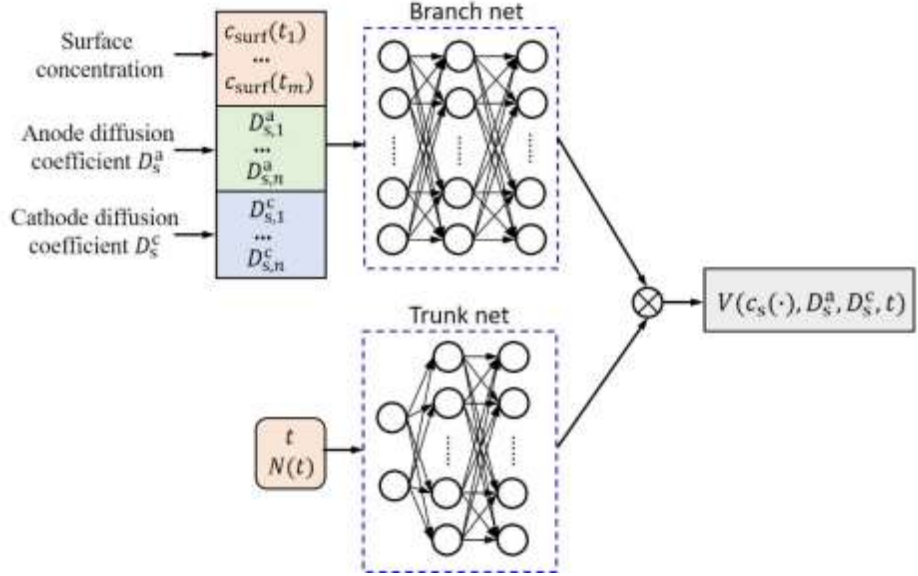

**Fig. 11.** DeepONet for learning the operator from Li-ion concentration at the surface of electrode particles to terminal voltage. The input surface concentration function is concatenated with two vectors filled with equal values of diffusion coefficients from anode and cathode, respectively.

Based on the trained DeepONet, terminal voltage can be predicted efficiently with new surface concentration and diffusion coefficients of two electrodes. The reference data from the two scenarios used in the last subsection are still used here. As shown in Fig. 12, the predicted voltage matches well with their reference values, with the maximum deviation for scenario 1 of approximately 7 mV and that for scenario 2 of approximately 9 mV. Even though we do not adopt PI-DeepONet here due to interpolated operations in underlying physics, we believe that as long as the Li-ion concentration is precisely predicted via PI-DeepONet in the last subsection, it is highly possible to obtain accurate predictions of terminal voltage, since surface concentration plays a key role in determining terminal voltage. Through the experiments, we can confirm that the complex mapping from current to terminal voltage considering the effects of solid diffusivity can be perfectly approximated by two consecutive surrogates built upon operator networks. Based on the composite surrogates, not only can forward mapping be efficiently realized, but inverse modeling can also be conducted, which will be discussed in the next section.

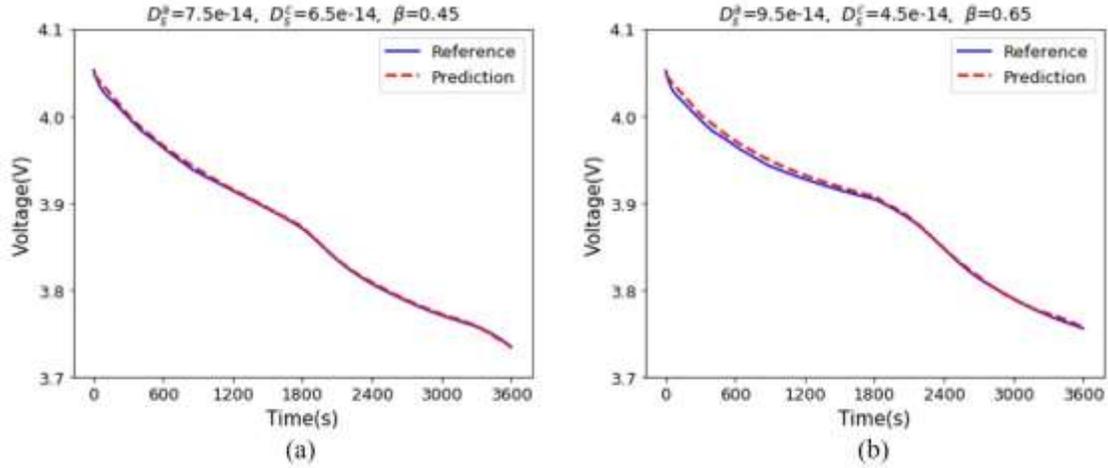

**Fig. 12.** The matching of predicted terminal voltages and their references, with (a) corresponding to scenario 1 and (b) corresponding to scenario 2.

**3.3 Parameter estimation based on composite surrogate model**

With the utilization and degradation of batteries, certain material parameters may evolve with time, and thus they need to be updated according to real-time measurements when adopting them for forward simulations. Since the original physics-based model cannot combine with data naturally, it is required to employ optimization methods (e.g., genetic algorithm [43]) or Bayesian methods (e.g., Kalman filter [44] and Markov Chain Monte Carlo [45]) for parameter estimation, leading to repeated forward simulations that are time-intensive. As a solution, constructing a surrogate for the physics-based model to accelerate parameter estimation, especially for on-board applications, such as BMS, is quite necessary. Apart from computational efficiency, the surrogate built with operator networks is differentiable anywhere, which endows it with the ability to embrace data directly and update input parameters via gradient-based optimization. Therefore, in this work, we build a totally differentiable surrogate system, as shown in Fig. 13, which is composed of two trained PI-DeepONets used for predicting Li-ion concentration based on current and diffusion coefficients at two electrodes, and one trained DeepONet used for predicting terminal voltage based on surface concentration predictions and diffusion coefficients. As long as the measurements of terminal voltage are obtained, the input parameters can be updated.

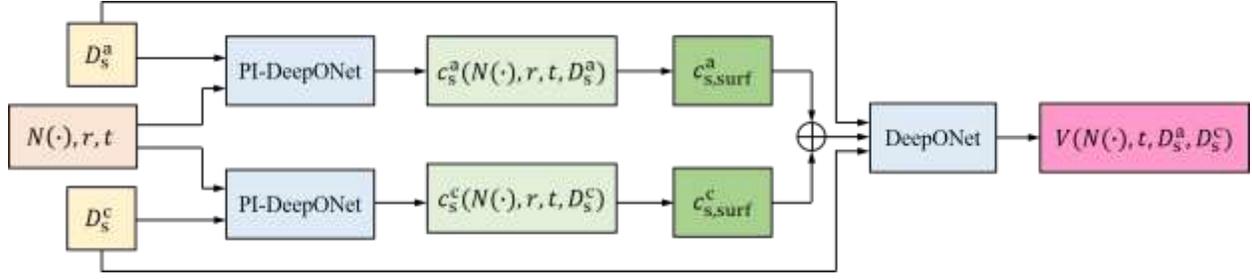

**Fig. 13.** Schematic of a composite surrogate, which is composed of two parallel PI-DeepONets and one subsequent DeepONet. The two PI-DeepONets are utilized to predict Li-ion concentration of anode and cathode, respectively, based on the shared current function and their corresponding solid diffusion coefficients. Upon obtaining the concentration in two electrodes, the third DeepONet can leverage their surface values to predict terminal voltage by incorporating the diffusion coefficients simultaneously.

In the experiment of this work, we only consider three changeable parameters, i.e., diffusion coefficient in anode $D_s^a$, diffusion coefficient in cathode $D_s^c$, and $\beta$ that determines current function, in model inputs. Prior to conducting parameter estimation, we firstly test their sensitivity on terminal voltage. The ranges for the three parameters are set as $D_s^a$ ($D_s^c$) $\in [2e-14, 1e-13]$ and $\beta \in (0,1)$. We randomly select 30 samples for one parameter, while keeping the other two as median values of their ranges, and calculate the corresponding terminal voltages. As shown in Fig. 14, the variations in $\beta$ impart the most remarkable effects on voltage (Fig. 14(c)), while those in $D_s^a$ have relatively weaker influences (Fig. 14(a)). Meanwhile, almost no effects of $D_s^c$ on voltage can be observed from Fig. 14(b), possibly attributable to a narrow parameter range. Therefore, in this case, only $D_s^a$ and $\beta$ are suggested to be inferred from measurements of terminal voltage.

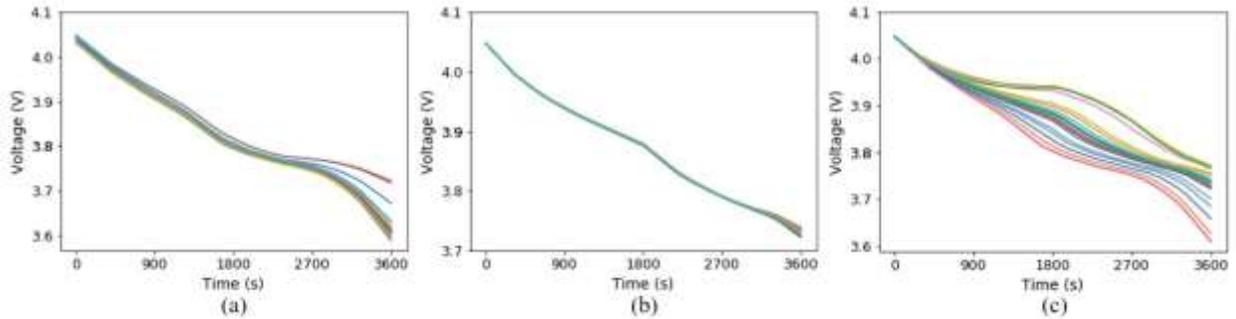

**Fig. 14.** Sensitivity tests of (a) anode diffusion coefficient $D_s^a$, (b) cathode diffusion coefficient $D_s^c$, and (c) $\beta$ value, on terminal voltage.

The performance of parameter inference is tested with two scenarios, which differ in the standard deviation of measurement error, with 5 mV for scenario 1 and 10 mV for scenario 2. In practice, we assign the parameters to be inferred with true values, and use them to generate synthetic observations with measurement noise being injected. Based on the trained composite surrogate, the input parameters to be estimated, i.e., $D_s^a$ and $\beta$, are treated as variables and optimized using MSE loss through the Adam optimizer with a learning rate of 0.01. For each scenario, two parallel tests with different assumed true values are conducted, and each test, which takes approximately 42 s for 100 iterative operations, is repeated 30 times with different initializations to reduce randomness. The boxplots for 30 initial and finally estimated values in the two scenarios are given in Fig. 15, with blue triangles representing the assumptive true values. It can be seen from Fig. 15(a) that $\beta$ can be accurately inferred from voltage measurements in the two scenarios. In comparison, as shown in Fig. 15(b), $D_s^a$ is estimated with relatively larger uncertainties, since terminal voltage is less sensitive to $D_s^a$ than $\beta$, and $D_s^a$ bears larger risks from ill-posedness. However, it is fortunate that the true values of $D_s^a$ can be covered narrowly by their estimated regions, and thus the estimation accuracy is still acceptable.

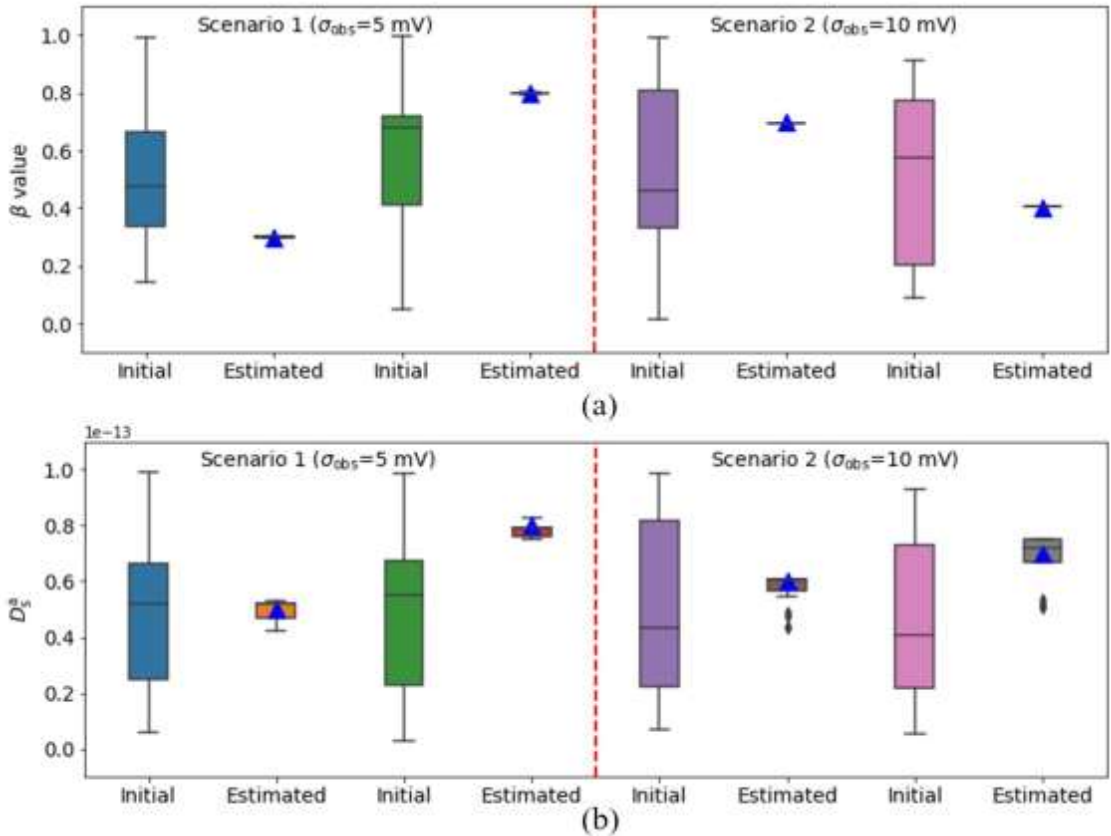

**Fig. 15.** Estimation for (a) $\beta$ value and (b) anode diffusion coefficient $D_s^a$ in the two scenarios, which differ in the level of measurement noise. In each scenario, two parallel reference values are considered, which are denoted by blue triangles.

## 4. Conclusion

Motivated by the necessity of developing a battery model with both accuracy and efficiency, in this work, we firstly treated the functional mapping from current function to terminal voltage as a composite of operators, which can be approximated by data-driven methods while incorporating the underlying physics as constraints.

Specifically, we adopted a recently popular method, termed DeepONet, for operator learning. To validate its learning capability, we built an operator to map from current curve to Li-ion concentration at two electrodes, and found that the physics-informed DeepONet possessed great potential to outperform the purely data-driven DeepONet, especially in temporal extrapolation scenarios. Subsequently, we continued to construct a composite surrogate for mapping current to voltage with solid diffusivity being considered using three operator networks. Two parallel physics-informed DeepONets were firstly utilized to predict Li-ion concentrations at two electrodes based on current function and solid diffusion coefficients. Then, based on the predicted concentrations at the surface of electrode particles, a DeepONet was built to map it to terminal voltage with solid diffusivity also being considered. Therefore, the composite surrogate was built upon the trained models by taking Li-ion concentration as an intermediate quantity, which can be adopted for calculating SOC used in BMS. Based on the surrogate, which is differentiable anywhere, its natural capability for combining data was validated through an inverse test, i.e., using measurements of terminal voltage to estimate input parameters accurately and efficiently.

Even though the electrochemical performance, i.e., Li-ion concentration and terminal voltage, and input parameters, i.e., diffusion coefficient in anode and $\beta$ that determines current, can be inferred with satisfactory accuracy and efficiency through operator networks, certain shortcomings should be addressed in future work. Firstly, based on the trial work in which only the discharge scenario was considered, we will conduct experiments in scenarios with both charge and discharge, which is closer to reality. In addition, the underlying physics utilized in this work are a relatively simplified model, i.e., the eSP model, which may lose reliability in high charge and discharge rates. In the future, the P2D model with the highest fidelity will be used for building operator networks, with an emphasis on imposing constraints from multiphysics, for which the pioneering works [46,47] can be referred.

## Acknowledgments

This work is partially funded by the Shenzhen Key Laboratory of Natural Gas Hydrates (Grant No. ZDSYS20200421111201738), the SUSTech - Qingdao New Energy Technology Research Institute, and the China Postdoctoral Science Foundation (Grant No. 2020M682830).

## Appendix

## A. Coefficients in the polynomial expressions for Li-ion concentration in electrolyte phase

The coefficients $a_k$ and $b_k$ used in polynomial expressions for approximating Li-ion concentration in the electrolyte phase are given by [38]:

$$a_1 = -\frac{\varepsilon_n^{-bru} J}{2 D_e L_n}, \tag{A.1}$$

$$a_2 = J[\varepsilon_n^{1-bru} L_n^2 + 2\varepsilon_p^{1-bru} L_p^2 + 6\varepsilon_p \varepsilon_{sep}^{-bru} L_s L_p + 3\varepsilon_{sep}^{1-bru} L_{sep}^2 + 3\varepsilon_n^{-bru} L_n (\varepsilon_p L_p + \varepsilon_{sep} L_{sep})]/6 D_e (\varepsilon_n L_n + \varepsilon_p L_p + \varepsilon_{sep} L_{sep}) \tag{A.2}$$

$$a_3 = -\frac{\varepsilon_p^{-bru} J}{2 D_e L_p}, \tag{A.3}$$

$$a_4 = -J[\varepsilon_p^{1-bru} L_p^2 + 2\varepsilon_n^{1-bru} L_n^2 + 6\varepsilon_n \varepsilon_{sep}^{-bru} L_n L_p + 3\varepsilon_{sep}^{1-bru} L_{sep}^2 + 3\varepsilon_n^{-bru} L_p (\varepsilon_n L_n + \varepsilon_{sep} L_{sep})]/6 D_e (\varepsilon_n L_n + \varepsilon_p L_p + \varepsilon_{sep} L_{sep}) \tag{A.4}$$

$$b_1 = -6 D_e \varepsilon_n^{bru-1} \varepsilon_p^{bru} \varepsilon_{sep}^{bru} (\varepsilon_n L_n + \varepsilon_p L_p + \varepsilon_{sep} L_{sep})/L_n [2\varepsilon_n^{bru} \varepsilon_p \varepsilon_{sep}^{bru} L_p^2 + 2\varepsilon_p^{1+bru} L_p (\varepsilon_{sep}^{bru} L_n + 3\varepsilon_n^{bru} L_{sep}) + \varepsilon_p^{bru} \varepsilon_{sep} L_{sep} (2\varepsilon_{sep}^{bru} L_n + 3\varepsilon_n^{bru} L_{sep})] \tag{A.5}$$

$$b_2 = -6 D_e \varepsilon_p^{bru-1} \varepsilon_n^{bru} \varepsilon_{sep}^{bru} (\varepsilon_n L_n + \varepsilon_p L_p + \varepsilon_{sep} L_{sep})/L_p [2\varepsilon_n \varepsilon_p^{bru} \varepsilon_{sep}^{bru} L_n^2 + 2\varepsilon_n^{1+bru} L_n (\varepsilon_{sep}^{bru} L_p + 3\varepsilon_p^{bru} L_{sep}) + \varepsilon_n^{bru} \varepsilon_{sep} L_{sep} (2\varepsilon_{sep}^{bru} L_p + 3\varepsilon_p^{bru} L_{sep})] \tag{A.6}$$

where $bru = 1.5$ is the Bruggeman coefficient; $J = (1 - t_+) \frac{i_{app}}{F}$, in which $t_+$ is the Li-ion transference number and set as 0.363, and $F$ is the Faraday constant and set as 96485.33; and $D_e = 3 \times 10^{-10}$ is the diffusion coefficient in the electrolyte phase.